# Quantum logic as reversible computing

Dr. Basil Evangelidis
PhD University of Athens
FernUniversität in Hagen
Email : vasevang@ieee.org

*The relation between entropy and information has great significance for computation. Based on the strict reversibility of the laws of microphysics, Landauer (1961), Bennett (1973), Priese (1976), Fredkin and Toffoli (1982), Feynman (1985) and others envisioned a reversible computer that cannot allow any ambiguity in backward steps of a calculation. It is this backward capacity that makes reversible computing radically different from ordinary, irreversible computing. The proposal aims at a higher kind of computer that would give the actual output of a computation together with the original input, with the absence of a minimum energy requirement. Hence, information retrievability and energy efficiency due to diminished heat dissipation are the exquisite tasks of quantum computer technology.*

In the beginning of the twenty first century researchers and developers in microphysics try to guess whether solar fusion or quantum computing is going to be first successfully applied. An important application of featured quantum computing is quantum simulation. Complicated phenomena, such as the relation of the sensitivity of cells with the quantum principle of superposition, disclose the significance of the project for the construction of a universal quantum computer (Deutsch, 1985; Brown, 1999). The basic idea is to make use of quantum parallelism, according to which two quite different things must be considered as taking place simultaneously in quantum linear superposition – like the photon simultaneously being reflected and passing through the half-silvered mirror, or passing through each of two slits. With a quantum computer, these two different but superposed things would, instead, be two different computations (Penrose, 1989, p. 518). The model of the quantum computer seems better apprehensible through the quantum theory of universal wavefunction (Everett, 1956), which later took the form of the many worlds interpretation of quantum theory, epitomized by DeWitt (1970, p. 30) with the following sentence: "Could the solution to the dilemma of indeterminism be a universe in which all possible outcomes of an experiment actually occur?" Under this perspective, the quantum computer as an advanced project of theorizing is a demand for the discovery of the best possible explanations, as Deutsch (1998) proposes.

In 1900, Hilbert motivated all future Quantum Theory by communicating the following request: "To find a few physical axioms that, similar to the axioms of geometry, can describe a theory for a class of physical events that is as large as possible." This requirement to axiomatize physics was a direct consequence of the probabilistic results of the augmentation of entropy in thermodynamics. Feynman would later exemplify this consequence by the mental experiment of a gas with N atoms (or molecules), occupying a volume V, taken firstly with the supposition of no forces of attraction or repulsion between each constituent. If we applied slow, isothermal compression to this gas, allowing thus simultaneously for draining off the gained internal heat through a surrounding thermal bath, the occupied volume of the gas would decrease due to the compression. The total energy U of the gas, however, would remain stable, because the process would be isothermal. In addition, our knowledge of the possible locations of molecules would increase with the decrease of the occupied volume. If the probability of a particular gas configuration is W, we have entropy: $S = k \log W$. The bigger W, the bigger the randomness and the disorder, the bigger the entropy, the more states it could be in, the less we know about the ensemble of the configurations. Since information is certainly considered as the opposite of disorder, entropy can be used to characterize the information content in a signal.

We need a distinct quantum logic, for significant reasons. Since microphysical research assumes the existence of unobservable entities, theory of knowledge adopts beliefs taken from different standpoints, such as the instrumentalist, the logical positivist, the logical empiricist, the scientific realist, the internal realist, the constructive empiricist and the semantic conception about the reality of unobservables. The development of the above mentioned standpoints was a practical deepening of previous physical philosophies, on the basis of evidence from quantum phenomena. At this high level of theoretical diversity, the correspondence between theory and phenomena is mediated by models. From ancient atomism to modern quantum theory, the main argument in favour of the existence of unobservables was based on mathematical realism, pragmatism and optimism. With thermodynamics and quantum theory, scientific realism becomes also amenable to the probability theory of meaning. Scientific and structural realism, however, is criticised by scientists who propose a philosophy of structural empiricism, which argues that empirical confirmation is limited to finite quanta due to line spectra. According to Born (1956), the Gordian knot was cut by Heisenberg (1925) with a) the philosophical principle on the restriction of theoretical description to observable facts, and with b) the mathematical introduction of the matrix mechanics,[1] that is to say, square arrays, which diagonally represent states and non-diagonally transitions, instead of unobservable electron orbits with definite radii and periods of rotation. The motion thus was not represented by a coordinate as a function of time, but by an array of transition amplitudes:

$$\begin{array}{cccc} 11 & 12 & 13 & - \\ 21 & 22 & 23 & - \\ 31 & 32 & 33 & \end{array}$$

Positions on a diagonal correspond to states, and non-diagonal positions correspond to transitions. Despite empiricism, the logical exploitation of quantum theory does not cut its

---

[1] Matrix mechanics is a formulation of quantum mechanics created by Werner Heisenberg, Max Born, and Pascual Jordan in 1925. It was the first conceptually autonomous and logically consistent formulation of quantum mechanics. Its account of quantum jumps supplanted the Bohr model's electron orbits.

ties with mathematical applicability. Let's consider a conventional logic gate, such as the implication:

$$p \rightarrow q$$

$$q \rightarrow r$$

$$\overline{\phantom{aaaaaaaa}}$$

$$p \rightarrow r$$

for instance,

$$\text{perceive } x \rightarrow \text{name } x$$

$$\text{name } x \rightarrow \text{learn } x$$

$$\overline{\phantom{aaaaaaaaaaaaaaaaaaaa}}$$

$$\text{perceive } x \rightarrow \text{learn } x$$

The truth table of the logic gate of implication can clearly exhibit that only if the output is F can we know with certainty that the input is T → F, therefore, the possibility to avoid loss of energy and information (increase of entropy) by knowing the input from the output is only 0,25:

| input | input | output | entropy |
|-------|-------|--------|---------|
| T | T | T | ↑S |
| T | F | F | ∴ |
| F | T | T | ↑S |
| F | F | T | ↑S |

Take a black box with a set of input and output lines. Suppose that for any input there is one and only one output, and that this is determined by the input. In the most trivial case, the signals simply propagate through the box unchanged. Hereby the computation is reversible, since the output carries no more information than the input. Information in this case is retrievable. By contrast, all conventional logic gates, with the only exception of NOT, lose information irretrievably.

The proposal aims at a higher kind of computer that would give the actual output of a computation together with the original input, with the absence of a minimum energy requirement. Information retrievability and energy efficiency due to diminished heat dissipation are the exquisite tasks of quantum computer technology. This project is compatible with the ideas of the philosopher of technology Simondon (2006; 2009), who stressed that the tools and the machines correspond to information and energy resources. Thus, the projects for quantum reversible computing, based on the relation between information and entropy, promise an energy saving, reversible computing:

> *The reversibility of physics means that we can never truly erase information in a computer. Whenever we overwrite a bit of information with a new value, the previous information may be lost for all practical purposes, but it hasn't really been physically destroyed. Instead it has been pushed out into the machine's thermal environment, where it becomes entropy -in essence, randomized information- and manifests as heat* (Frank, 2017).

In 1961, Rolf Landauer and John Swanson at IBM began researching information dissipation, heat generation and reversible computing. Landauer's younger colleague, Charles Bennett showed in 1973 that it is possible to construct fully reversible computers capable of performing any computation without erasing information and without stuffing up memory with temporary data. In fact, any logic gate that has more input than output lines inevitably loses information, because we cannot figure out the input from the output.

The Fredkin gate and the Toffoli gate, which were constructed in the late 1970s and early 1980s at M.I.T., were the first investigations on reversible computing, where the input can always be deduced from the output. The Fredkin gate has three input lines and three outputs. The one line is called the control channel. If the control channel is set at 0, the other two lines are also unchanged. If the control line is at 1, however, the outputs of the other two lines are switched. Furthermore, after Paul Benioff and Richard Feynman, information processing is conceived as a simulation of the values of the spin[2] of a particle to one of the outputs of a logic gate (Bennett and Landauer, 1985).

As Otto Stern and Walther Gerlach had found in 1921, a magnetic field deflects particles along only two paths, indicating that the spin is quantized. However, experiments showed that it is impossible to simultaneously assign exact values to the x and y components of the spin of a particle. Particles with spin 1/2 under a $2\pi$ rotation transform as $\psi \rightarrow -\psi$ because they cannot be smoothly deformed to a point. This experimental data is persuading enough that quantum field theory cannot conform to classical logic.

Reversible computing is feasible by running back to the input the intermediate calculations of the computational output and reversing the computer to its initial state. Dr. Bennett likened the method to crossing a river using just a few stepping stones: one must backtrack to pick up the stones left behind, placing them ahead. In 1993, Saed Younis in Tom

---

[2] According to the spin-statistics theorem in relativistic quantum field theory, particles with integer spin are bosons, while particles with half-integer spin are fermions. A fermion is a particle that follows Fermi–Dirac statistics and generally has half odd integer spin: spin 1/2, spin 3/2, [contradictory] etc. These particles obey the Pauli exclusion principle. Fermions include all quarks and leptons, as well as all composite particles made of an odd number of these, such as all baryons and many atoms and nuclei. Fermions differ from bosons, which obey Bose–Einstein statistics.

Knight's group showed that adiabatic circuits could be used to implement fully reversible logic.

Quantum mechanics differ from classical, as their "dynamical variables do not obey the commutative law of multiplication," since position and momentum are conjugate variables. These dynamic variables are not ordinary c-numbers but what Dirac calls "q-numbers", representable "by matrices whose elements are c-numbers (functions of a time parameter)" (Dirac, 1927, p. 621).[3] In 1925, Heisenberg informed Born about his discovery of the matrices, as Born recalls:

> *The significance of the idea was at once clear to me and I sent the manuscript to the* Zeitschrift für Physik. *I could not take my mind off Heisenberg's multiplication rule, and after a week of intensive thought and trial I suddenly remembered an algebraic theory which I had learned from my teacher, Professor Rosanes, in Breslau. Such square arrays are well known to mathematicians and, in conjunction with a specific rule for multiplication, are called matrices. I applied this rule to Heisenberg's quantum condition and found that this agreed in the diagonal terms. It was easy to guess what the remaining quantities must be, namely, zero; and at once there stood before me the peculiar formula*
>
> $$pq - qp = h/2\pi i$$
>
> *This meant that coordinates q and momenta p cannot be represented by figure values but by symbols, the product of which depends upon the order of multiplication - they are said to be «non-commuting»* (Born, 1954, p. 259).

The q-numbers are the foregoers of the qubits. Quantum information is reducible to qubits, to one- and two-qubit gate operations. Qubits are the quantum counterparts of the binary bits. Their physical counterparts are polarized photons. A qubit can have a value that is either 0, 1 or a quantum superposition of 0 and 1. A two-dimensional column vector of real or complex numbers represents a possible quantum state held by a qubit, represented as $\begin{bmatrix}\alpha\\\beta\end{bmatrix}$ that is a qubit state with the complex numbers α and β satisfying

$$|\alpha|^2 + |\beta|^2 = 1.$$

Here are some examples of valid quantum state vectors representing qubits are:

---

[3] Heisenberg had already developed diagonal matrices that their elements represented either the energy levels of the system, or the total polarization, as periodic functions of time that determine the frequencies and intensities of the spectral lines. "Further developments of the theory of exchange have been made by Heitler, London and Heisenberg, containing applications to molecules held together by homopolar valency bonds and to ferromagnetism. The treatment given by these authors makes an extensive use of group theory and requires the reader to be acquainted with this branch of pure mathematics. Now group theory is just a theory of certain quantities that do not satisfy the commutative law of multiplication, and should thus form a part of quantum mechanics, which is the general theory of all quantities that do not satisfy the commutative law of multiplication" (Dirac, 1929, p. 716).

$$\begin{bmatrix}1\\0\end{bmatrix}, \begin{bmatrix}0\\1\end{bmatrix}, \begin{bmatrix}\frac{1}{\sqrt{2}}\\\frac{1}{\sqrt{2}}\end{bmatrix}, \begin{bmatrix}\frac{1}{\sqrt{2}}\\\frac{-1}{\sqrt{2}}\end{bmatrix}, \begin{bmatrix}\frac{1}{\sqrt{2}}\\\frac{i}{\sqrt{2}}\end{bmatrix}$$

The decoherence of quantum superpositions through the interaction of the computer with the environment, and the precise application of quantum state transformations in order to obtain accurate results after many computation steps, are considered as the main difficulties of the quantum computer project (Shor, 1997). According to the superposition principle, between any two reliably distinguishable states of a physical system, such as vertically and horizontally polarized single photons, there are intermediate states that are not reliably distinguishable, for instance diagonal photons. If photons pass through polarization filters, they cannot amount to irrational or rational numbers, but only to natural numbers. For this reason, their quantity is substituted by probability (Laidler, 1998).

A photon passing through an interferometer is considered, then, as being in a translational state, resulting by the superposition of two probable states, until with a quantum jump interferes with itself. However, these general phenomena are not evident, because "for waves of familiar frequencies the associated quanta are extremely small, while for particles even as light as electrons the associated wave frequency is so high that it is not easy to demonstrate interference" (Dirac, 1967, p. 10). In general, the principle of superposition that occurs in quantum mechanics is intimately interconnected with a physical interpretation that presupposes indeterminacy in the results of observations. A superposition of inputs gives a superposition of outputs, which equals to an entangled state, based on the exact sameness of polarization, like a pair of magic coins.

The concept of a quantum state assigns a probability distribution to the outcomes of each possible measurement on a system. The behaviour of a quantum system is determined by the quantum state together with the rules for the system's evolution. A pure quantum state corresponds to a ray in a Hilbert space over the complex numbers, while mixed states are represented by density matrices. Individual qubits making up a multi-qubit system, cannot always be characterized as having individual states of their own. More precisely, they do not always have what are called pure states of their own. Hence, we give a statistical description of an individual qubit, or a group of qubits, in terms of a density matrix or mixed state (Mermin, 2007).

### Quantum logic gates

In 1925, Heisenberg had argued that research in microphysics must be carried out not according to orbits or times of rotation of electrons in the atom, but rather upon measurable differences in the radiation frequencies and spectral line intensities, in order to focus solely on the following purpose: "To develop a quantum-theoretical mechanics analogous to classical mechanics, in which only relationships between observable quantities occur." The ultra-microscopic object, a corpuscle, taken in its physical role, it is rather a means of analysis that an object of empirical knowledge. It is a pretext of thought, as it seems that in the world of microphysics the unique loses its substantial properties. There are, hence, substantial properties only above - not below - microscopic objects, as Bachelard (1931-32) remarked. Any attempt to consider quantum theory in terms of substantiality would equal to an *epistemological obstacle*, which hinders theoretical advance. In the beginning of the mathematical conception of the quantum lays the relation, which permits mathematics reign over microphysical reality.

The quantum principles of complementarity, uncertainty, and non-commutativity of observations, do not conform to classical logic and let novel logical and epistemological notions emerge, like the semantic conception of theories (Birkhoff & von Neumann, 1936; Rédei, 2014; Summers, 2016). Quantum theory and experiment assert either the existence of an increasing population in the particle zoo, or the substitution of ontology for virtual densities and entangled affinities. Entanglement is a pure quantum feature that is absent from classical systems. A state is entangled with respect to the disjunction of two sub-algebras iff it is not a product state. If the sub-algebras are quantum (non-abelian) then pure states can be entangled, and no one doubts that these states are examples of quantum entanglement. When separable and independent quantum systems become entangled through temporary physical interaction, they can no longer be described as before, since their $\psi$-functions have been entangled, as Schrödinger (1935, p. 555) suggested, regarding it as the most significant characteristic of QM.

Schrödinger supported a proposal to dispense with the particle representation entirely, and instead of speaking of electrons as particles, to consider them as a continuous density distribution $|\psi^2|$ (or electric density $e|\psi^2|$). In fact, Schrödinger did not believe in particles, because "in general, they do not incorporate coexistent virtualities, they cannot be reidentified through time, and they do not play the role of individual substances bearing properties" (Bitbol, 2007: p. 89). Schrödinger believed that the electrons are wave-matter, "wave packets," "wavicles" rather than discrete particles. It is no coincidence that the primary influence on Schrödinger was de Broglie. On the contrary, ontological commitment in microphysics is expressed by the Kantian interpretation of quantum mechanical complementarity, with its emphasis on the classical entity of the wave (Bohr) and the introduction of a multiple quantity of sub-particles, such as the quarks.[4]

> *Only five years ago* [1959] *it was possible to draw up a tidy list of 30 subatomic particles that could be called, without too many misgivings, elementary. Since then another 60 or 70 subatomic objects have been discovered, and it has become obvious that the adjective "elementary" cannot be applied to all of them...*
> 
> *... early in this century... spectroscopists, studying the light emitted by excited atoms, found a profusion of discrete wavelengths that were at total variance with the wavelengths predicted by classical electrodynamics....*

---

[4] The previous scientific hypothesis that the positive charges of the protons should make them repel one another and dismantle the nucleus had never been observed. The binding energy of the nucleus, therefore, should be much stronger than formerly believed, acting on protons and neutrons. This binding energy is actually a consequence of the fundamental interaction that attracts the quarks together, expressed as colour charge and mediated through the gluons.

> *…. All particles discovered to date participate in this strong interaction except the photon (the particle of light and other electromagnetic radiation) and the four particles called leptons: the electron, the muon (or mu particle) and the two kinds of neutrino* (Chew et al., 1964, p. 74).

In quantum computation, the state of the computer is described by a state vector $\psi$. Quantum logic gates are represented by unitary matrices. Qubits obey a no-cloning rule, which also precludes sending a qubit state to two different gates at the same time. A quantum gate on one qubit is described by a 2×2 matrix, and a quantum gate on two qubits by a 4×4 matrix. A gate which acts on n qubits is represented by a $2^n \times 2^n$ unitary matrix. The quantum states that the gates act upon are vectors in $2^n$ complex dimensions. The state's evolution in the course of time t is described by a unitary operator U on this vector space, thus, a linear transformation which is bijective and length-preserving. The current notation for quantum gates was developed by Barenco et al. (1995), building on notation introduced by Feynman (1985).

> *Historically, the idea that the quantum mechanics of isolated systems should be studied as a new formal system for computation arose from the recognition twenty years ago that computation could be made reversible within the paradigm of classical physics. It is possible to perform any computation in a way that is reversible both logically—i.e., the computation is a sequence of bijective transformations—and thermodynamically—the computation could in principle be performed by a physical apparatus dissipating arbitrarily little energy* (Barenco et al., 1995).

> *… the time evolution of a classical reversible computer is described by unitary operators whose matrix elements are only zero or one — arbitrary complex numbers are not allowed… The quantum gate array is the natural quantum generalization of acyclic combinational logic "circuits" studied in conventional computational complexity theory* (Barenco et al., 1995)

The reversible operations that a quantum computer can perform upon a single qubit are represented by the action on the state of the qubit of any linear transformation that takes unit vectors into unit vectors. These transformations are called unitary and satisfy the condition $uu^\dagger = u^\dagger u = 1$. Since any unitary transformation has a unitary inverse, such actions of a quantum computer on a qubit are reversible. The only nontrivial reversible operation a classical computer can perform on a single qubit is the NOT operation X. The operations AND, NAND, OR and XOR are irreversible. Hence, from the output of the gate you cannot reconstruct the input: information is irreversibly lost. The functionality of the quantum logic gates, on the contrary, is reversible, since we can always infer the input from the output. For instance, the CONTROLLED NOT (CN) gate allows for reversible computing through information conservation in the control line and input inversion for the NOT line (only if the input in the control line is 1).

The CONTROLLED CONTROLLED NOT (CN) gate obtains two control lines for the conservation of passing signals. The NOT line is activated only if the input in both control lines is 1. In this case the CCN gate is a NOT. If only A=1, then the gate is just a CN gate.

## Quantum circuit model and unitary transformation

In the theory of Maxwell the electromagnetic field is represented by the composition of two vectors, the electric that acts on immobile and moving loads, and the magnetic that acts only upon currents, that is to say, only on moving loads. Every kind of radiation, such as the light and the hertz waves of X and γ rays, appears like the electromagnetic perturbations, which propagate freely in space. All kinds of radiation belong to this general notion of the field. Gravitation force, however, exhibits a character that conserved the conception of action at a distance a long time after Maxwell, since the electromagnetic conceptions of the field are not applicable to gravitation. Only after the introduction of general relativity could gravitation be conceived in terms of a gravitational field. According to de Broglie, the triumph of the notion of the field was exemplified by the opposition between the matter, formulated locally by elementary particles, and the field which is extended throughout the void spaces that separate the elementary particles. Nevertheless, this conception has been rejected after the appearance of wave mechanics. While photons are emitted and absorbed as units, they are geometrically represented by vectors and tensors and they obey the Bose-Einstein statistics, by contrast, the elementary particles of matter appear and disappear in pairs, they are geometrically represented by spinors and they obey the Fermi-Dirac statistics. These differences are fundamentally expressed by quantum quantity h/2π, which is called spin[5] (intrinsic rotation) and equals to 1/2 for the elementary particles, to 1 for photons and to 2 for gravitons. At even smaller distances, of less than $10^{-12}$ centimetres, and with greater force, due to the interior coherence of the atomic nucleus, takes place the interaction of the nuclear field (de Broglie, 1951).

By considering machine-independent computing, the Turing/Church Thesis, the Strong Church's Thesis, and the Invariance Thesis obtain a modified meaning: Any physically realizable computational device can be simulated by a Turing machine in a number of steps polynomial in the resources used by the computing device. A Turing machine is a mathematical abstraction of a digital computer, while quantum computers were until the end of the twentieth century hypothetical. Let the input to a quantum computer be classical information, which can be expressed as a binary string S of length k. We can encode this in the initial quantum state of the computer, which must be a vector in $\mathbb{C}^{2^n}$. Then we concatenate the bit string S with n-k 0's to obtain the length n string S0…0 and we initialize the quantum computer in the state $V_{S0…0}$. (Shor, 1997; 1998).

A dynamical system is reversible if from any point of its state set one can uniquely find a trajectory backward as well as forward in time. Thus, for a time-discrete system such as an automaton, its transition function is invertible, that is, bijective. Any operations (such as the clearing of a register) can be planned at the whole-circuit level rather than at the gate level, and most of the time can be replaced by an information-lossless variant (Toffoli, 1980). The reversibility of quantum computing is the outcome of the invertibility of the matrices. Rather than conventional logic gates, we work with quantum circuits. The circuit of the wires displays changes in atomic states, produced by the operation of a matrix.

---

[5] In the non-relativistic context, we represent the spin states as being up and down. In the relativistic limit, it is better to think of the electron as being polarized, just like a photon. Polarizations for 1/2 particles are usually called helicities. We distinguish between circular polarization (called left and right helicity) and linear polarization. "Linearly polarized electrons are like linearly polarized light, and the polarizations must be transverse to the direction of motion. So the electron moving in the z direction can either be polarized in the y direction or in the x direction… To get spin 1 from spin 1/2 and spin 1/2, the electron and positron have to be polarized in the same direction" (Schwartz, 2014, p. 86).

Since quantum signals are analogical and sensitive to noise, the main design task is to minimize the introduction of noise into the qubit, hence, the transmission of the qubit state should not pass through noisy long wires. For this reason, every part of the quantum circuit system should be in the same area. Qubits should also conform to a no-cloning rule, which also precludes sending a qubit state to two different gates at the same time.

A mathematical visualisation of the operations of quantum gates is attained by the representation of the state of 'n' qubits as a vector in a high dimensional space ($2^n$ complex dimensions). The value of the vector in each dimension is given by the complex coefficients $a_i$. The length of the vector remains constant and equal to 1, thus, the state of the system can be any place on the unit hypersphere, which is the outcome of the extension of a sphere to higher dimensions.

> *All quantum gates are simple rotations of the state vector to a new position on the hypersphere. As the number of qubits increases, the dimension of the space grows exponentially, but the state vector remains unit length, and the operations remain the different rotations possible on the hypersphere (which are all reversible). Operations that preserve the vector length are said to be "unitary"* (National Academies, 2019, pp. 42-43).

The quantum gates are designed to operate on inputs of one, two, or three qubits. Furthermore, a limited number of quantum gates can be used to create a universal gate set, which can approximate all possible quantum gate combinations. Measurement aims at the extraction of information from the quantum computer. However, measurements collapse the system wave function and return from the n-qubit quantum register only a classical result, namely, n bits of information.

## Is there a quantum-a-priori?

The Augustinian identity between being and truth was gradually transformed into the method of resemblance, a significant undertaking of intellectual life until the sixteenth century. It was a resemblance that formulated explanatory practices, as Foucault (2002) notes. Representation, repetition, similitude, were the organising principles of theoretical life. Foucault distinguished four main sorts of resemblance in the Middle Ages: a) *convenientia*, as local proximity, entanglement of place and similitude, b) *aemulatio*, a contactless influence at a distance, as the human face emulates the sky; a universal imitation and reflection, c) *analogy*, either between a set and its members, or between different species, for instance, animal and plant, man and bird, etc., and d) *sympathies*, varieties of attraction, such as between the sunflower and the sun; processes of assimilation and transformation. Sympathy is counterbalanced by its opposite, antipathy. The pairs of sympathy and antipathy are the most fundamental resemblances, playing the role of the explanans of the other three sorts of similitude.

The synthetic a-priori knowledge of science and mathematics is precondition for the building of technology. A mathematical truth is given by intuition, according to Poincaré. A principle is a priori synthetic, if it is not a logical one and if it gives mathematical knowledge, as is the case for induction. This is why, no reduction of mathematics to logic is possible. Therefore, mathematical objects exist in a different manner from logical entities; by analogy, technical objects could be unaccustomed to logical ones. What does the word "exist" in mathematics mean? It means, as Poincaré (1905) alleged, to be free from contradiction. This idea was disputed by Couturat: Logical existence, as he said, is quite another thing from the absence of contradiction. It consists in the fact that a class is not empty; to say: "There are α," it is, by definition, to affirm that the class α is not zero (Lolli 2013).

The question is, however, the following: Can quantum science be regarded as synthetic a-priori? Is there a quantum-a-priori?